\documentclass{article}

\usepackage{arxiv}
\usepackage[style=numeric,sorting=none, doi=false]{biblatex}
\usepackage{authblk}
\usepackage[svgnames]{xcolor}
\usepackage{graphicx}
\graphicspath{{images/}}
\usepackage{tikz}
\usepackage{pgfplots}
\pgfplotsset{compat=1.13}
\usepackage{caption}
\usepackage{subcaption}
\usepackage{multirow}
\usepackage{array}
\usepackage{booktabs}
\usepackage{pdflscape}
\usepackage[T1]{fontenc}
\usepackage[latin1]{inputenc}
\usepackage[italian]{varioref}
\usepackage{lmodern}
\usepackage{url}
\usepackage{comment}
\usepackage{mleftright,xparse}
\usepackage{amssymb,amsmath,amsfonts,amsthm,bm,mathtools,cuted}
\usepackage[ruled,vlined,linesnumbered]{algorithm2e}
\usepackage{siunitx}

\newcommand{\mc}[1]{\mathcal{#1}}

\providecommand{\keywords}[1]
{
  \small	
  \textbf{\textit{Keywords---}} #1
}

\definecolor{ultramarine}{RGB}{0,32,106}

\title{Deep Reinforcement Learning for URLLC data management on top of scheduled eMBB traffic}
\author[1]{Fabio Saggese}
\author[2]{Luca Pasqualini}
\author[1]{Marco Moretti}
\author[2]{Andrea Abrardo}
\affil[1]{Department of Information Engineering\\ University of Pisa\\ Pisa, 56122 Italy}
\affil[2]{Department of Information Engineering and Mathematics\\ University of Siena\\ Siena, 53100 Italy}
\affil[ ]{\texttt{fabio.saggese@phd.unipi.it, pasqualini@diism.unisi.it, marco.moretti@unipi.it, abrardo@dii.unisi.it}}

\begin{document}
\maketitle

\begin{abstract}
With the advent of 5G and the research into beyond 5G (B5G) networks, a novel and very relevant research issue is how to manage the coexistence of different types of traffic, each with very stringent but completely different requirements.  
In this paper we propose a \emph{deep reinforcement learning} (DRL) algorithm  to \emph{slice} the available physical layer resources between ultra-reliable low-latency communications (URLLC) and enhanced Mobile BroadBand (eMBB) traffic. 
 Specifically, in our setting the time-frequency resource grid is fully occupied by eMBB traffic and we train the DRL agent to employ proximal policy optimization (PPO), a state-of-the-art DRL algorithm, to dynamically allocate the incoming URLLC traffic by puncturing eMBB codewords.  
Assuming that each eMBB codeword can tolerate a certain limited amount of puncturing beyond which is in outage, we show that the policy devised by the DRL agent never violates the latency requirement of URLLC traffic and, at the same time, manages to keep the number of  eMBB  codewords in outage at minimum levels, when compared to other state-of-the-art schemes.
\end{abstract}

\section{Introduction}
\label{sec:introduction}
Resource slicing of different kinds of traffic is a key enabler for 5G and B5G  networks, allowing the coexistence on a common infrastructure of different services with different requirements such as eMBB and URLLC~\cite{Elayoubi2019}. 
The two kind of traffic have different quality-of-service (QoS): eMBB users require high throughputs, while URLLC has  strict low-latency and reliability constraints~\cite{Popovski2018}. In particular, URLLC traffic is characterized by short packets that need to be transmitted and decoded in less than 1 ms~\cite{She2017}, so that conventional channel-aware scheduling is generally not possible.

Addressing the problem of URLLC-eMBB scheduling, \cite{Popovski2018} compares the performance of different techiques in the uplink of a 5G system and lays the ground for the subsequent literature using either  \emph{puncturing}, orthogonal multiple access (OMA) and non-orthogonal multiple access (NOMA).  Immediate scheduling of URLLC packets in combination with hybrid automatic repeat request (HARQ) is another approach investigated in \cite{Anand2018}. In~\cite{Anand2020} eMBB codewords are punctured  to accomodate  URLLC traffic and the  throughput loss for eMBB packets is evaluated under different models.
In~\cite{Alsenwi2019}, the authors describe the process of resource allocation of eMBB-URLLC traffic employing an optimization problem taking into account the probability of presentation of an URLLC packet.
In~\cite{Tang2019} the authors propose a resource allocation scheme for URLLC-eMBB traffic based on  successive convex approximation and semidefinite relaxation of the general optimization problem. 

Because of its ability of finding very good to optimal policies for systems that dynamically change through time~\cite{sutton2018reinforcement}, \emph{reinforcement learning} is a natural choice to address the random dynamics of URLLC traffic.
Accordingly, in~\cite{Elsayed2019}, and~\cite{Li2019} the authors propose two RL algorithms based on Q-learning to multiplex eMBB and URLLC traffic employing OMA and NOMA, respectively. 
In~\cite{Huang2020}  DRL is employed to multiplex eMBB and URLLC traffic using  a deterministic policy gradient algorithm . 

Most of  the recent literature ~\cite{Popovski2018, Anand2020, Elsayed2019, Li2019, Huang2020} assumes that the URLLC packets are transmitted as soon as they arrive. However, within the URLLC latency a certain amount of delay can be tolerated so to give the scheduler some degree of freedom  to improve the performance of the system. Moreover,  in order to ensure slice isolation, the control planes of two different slices should be kept to a minimum degree of interaction~\cite{Anand2020}.

To minimize the impact on eMBB traffic, in this paper we address the slicing problem by allowing some slack for URLLC scheduling, letting the DRL agent choose when to transmit, within the latency constraint. To do so, the URLLC scheduler only needs to be informed about the robustness of each eMBB codeword to puncturing. Our proposed codeword model somehow resembles the \emph{threshold model} described in~\cite{Anand2020}, but it retains two important differences. First, we consider a more realistic non-homogeneous situation where different puncturing policies can be adopted at different times. The second difference is that we  consider a threshold per codeword rather than per user.
In the numerical results we will  show that our proposed scheduler is able to slice the resources to obtain good performance even if some eMBB codewords have no protection from puncturing.

\section{System Model}
\label{sec:system}
We consider a single cell scenario in which one base station (BS) serves a set of downlink user equipments (UE). 
The set of UEs belonging to the URLLC and eMBB slices are referred to as $\mc{E}$ and $\mc{U}$, respectively. We consider a single coherence interval as time horizon, where the channel can be considered constant. The time axis is divided into $\Sigma$ equally spaced time slots of fixed duration. To accomodate URLLC traffic, with its stringent latency requirements, slots are further divided into $M$ minislots\footnote{In 3GPP, the formal term for a ``slot'' is eMBB Transmit Time Interval (TTI), and a ``minislot'' is a URLLC TTI~\cite{Anand2020}.}.
As for the frequency domain, the system bandwidth is divided into $F$ orthogonal frequency resources (FR)\footnote{With ``frequency resources'' we refer to the abstract concept of bandwidth available in an OFDM system and we may refer to resource blocks or subcarriers, indifferently.}.

We consider two different schedulers, one for each type of traffic, which operate separately and independently of each other. The \emph{eMBB scheduler} is responsible for assigning time and frequency resources to eMBB users: each eMBB codeword can occupy any fraction of the total available number of minislots and FRs.
As customary, eMBB scheduling is operated at the slot boundaries.
At the same time, the \emph{URLLC agent} operates on a per minislot basis with the possibility of puncturing some of the resources already assigned to eMBB users, if needed. 
In the following, a detailed description of how we model eMBB traffic, URLLC traffic, and their coexistence is presented.

\begin{figure}[htb]
    \centering
    \includegraphics[width=8cm]{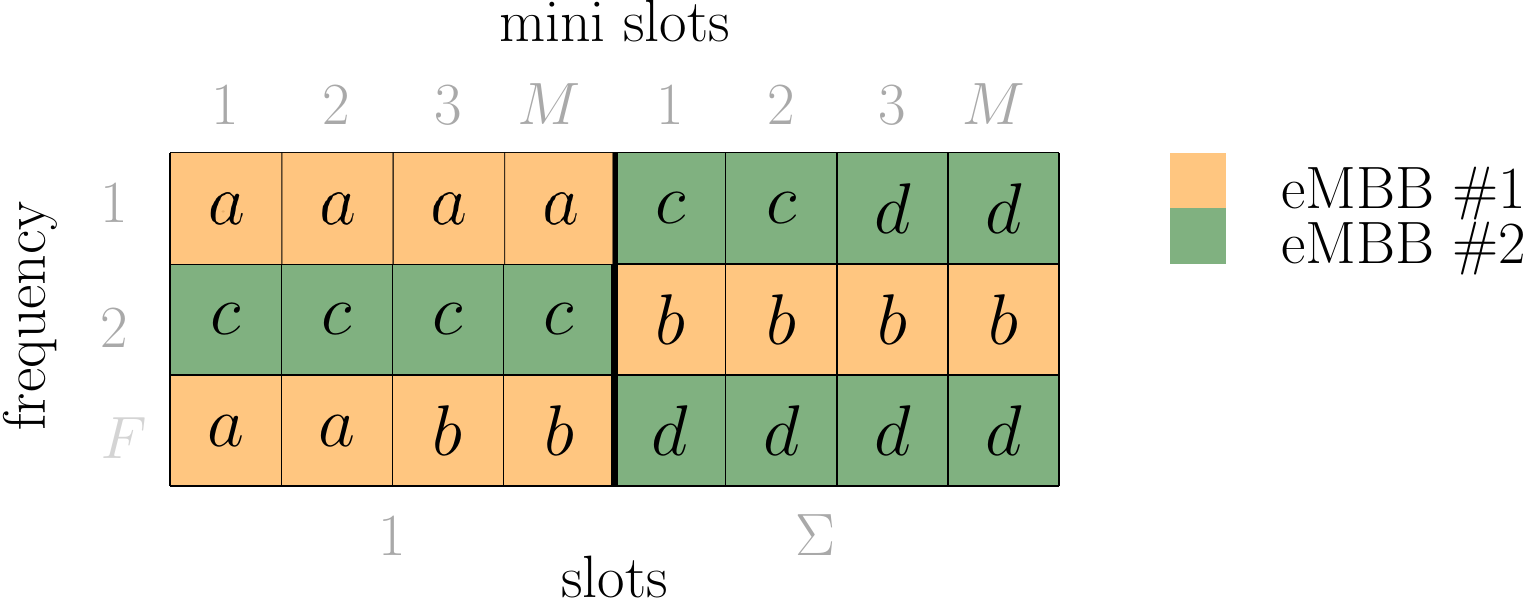}
    \caption{Toy example of the resource allocation and codeword placement for the eMBB users, $F=3$, $\Sigma = 2$, $M=4$. Resources are  allocated at slot boundaries, while codewords are $a,b\in\mc{W}_1$, $c,d\in\mc{W}_2$ and $|a|=|b|=|c|=|d|=6$.}
    \label{fig:ra}
\end{figure}

\subsection{The eMBB scheduler}
\label{sec:embb}
In this paper we do not explicitly address the eMBB scheduling problem, but, rather, we assume that a proper radio resource allocations has been performed somehow and we can focus on the coexistence of URLLC traffic on top of eMBB. Nevertheless, we need to describe the main principles of the eMBB scheduling policy, which is to maximize a rate-dependent utility function, not considering any latency. Hence, radio resources are allocated to the set of active users on a slot basis following the OMA paradigm. Moreover, since there is enough time to exchange channel quality information (CQI) before each scheduling decision, it is reasonable to assume perfect knowledge of  channel state information (CSI) at the BS. Therefore, eMBB resource allocation can be performed following conventional methods such as the water-filling algorithm~\cite{He2013}.

The scheduler has to further take into account that the eMBB packets might share the radio resources with URLLC traffic and in such event they should carry enough redundancy to be punctured without losing the entire packet.
We denote as $\mc{W}$ the codebook at the BS. The BS will then select a subset $\mc{W}_e \subset \mc{W}$ containing all the codewords of user $e$. A single codeword intended for user $e$ is denoted as $w \in \mc{W}_e$.
The length in symbols $|w|$ of a codeword is always a multiple of the minislot length, i.e., each codeword spans an integer number of minislots.
Finally, we denote with ${w}_{t,f}$ the codeword transmitted on the radio resource $f$ during minislot $t$  and with $\mc{W}_t=\bigcup_{\nu=1}^F {w}_{t,\nu}$ the set of all codewords transmitted during the minislot $t$.
Figure~\ref{fig:ra} shows a toy example of a possible resource allocation and codeword placement for two eMBB users.

\subsection{The URLLC DRL agent}
\label{sec:urllc}
Generally speaking, the QoS requirements of an URLLC user $u \in \mc{U}$ in a wireless network are specified as follows: a packet of size $N_u$ bits must be successfully delivered to the receiver within an \emph{end-to-end delay} of no more than $T_u^{\max}$ seconds with a probability of at least $1 - \epsilon_u$~\cite{Anand2018}. Moreover, a URLLC packet may randomly arrive at the BS at any moment. 
In this work we will concentrate on the edge delay, i.e. the delay computed as the difference between the time the scheduler receives the packet  and the time the packet is transmitted. This choice is justified by the fact that the backhaul delay is generally negligible~\cite{Rahman2018}, while UL queuing delay and transmission delay can be taken into account by reducing the value of the tolerable latency $T_u^{\max}$. Without loss of generality, we define the tolerable latency $T_u^{\max}$ in terms of the maximum number  $l_u^{\max}$ of minislots that can be waited before exceeding the latency constraint. 

To simplify the description of the problem, our work will focus on URLLC packets of fixed length corresponding to a single minislot. However, the considered framework can be easily extended to the case where a packet occupies more than one minislot. The packets are generated following a Bernoulli process, i.e.,  for each minislot there is a probability $p_u$ that a new URLLC  packet arrives. Then the packets are stored in a first-in first-out (FIFO) queue $\mc{Q}$ of infinite length.
The DRL URLLC agent is responsible for taking the  decision  whether the oldest packet in the queue should be transmitted or not in the current minislot.

Owing to the stringent latency constraint, the CSI of URLLC users cannot be estimated. Hence, power adaptation during transmission is not possible and ARQ re-transmission mechanisms can be hardly acceptable. Accordingly, reliability can be expressed in terms of outage probability for a fixed pre-defined transmit power.
As in previous works in literature~\cite{Anand2020}, we assume that the URLLC transmit power is large enough so that the outage probability remains under an acceptable threshold.

\subsection{URLLC and eMBB coexistence}
\label{sec:puncturing}
Coexistence of eMBB and URLLC is achieved by superposition coding or puncturing~\cite{Elayoubi2019}-\cite{Anand2020}. In this paper we consider a puncturing strategy, where the
BS decides to use a certain resource for URLLC traffic regardless of any eMBB user already occupying it. To avoid any interference between the two types of traffic, the eMBB codeword is punctured, i.e., the transmit power of the eMBB user on the  specific resource is set to zero.  
To tolerate puncturing, we assume that each eMBB codeword employs an inner erasure code with rate $1 - C_w/|w|$~\cite{Popovski2018}, that allows to correct up to $C_w$ erased minislots.
The eMBB scheduler is in charge of determining the \emph{class} $C_w$ for each codeword. The class assignment is performed on a codeword basis, i.e., the BS can assign codewords with different $C_w$ to the same user. The assignment of higher $C_w$ to different codewords encompasses the possibility of employing a more robust transmission mechanisms to prevent outage even in the presence of puncturing, as discussed in Section~\ref{sec:embb}.
Note that the algorithm implemented by the eMBB scheduler may be unknown by the URLLC traffic agent, as long as the latter is informed of the codewords allocation and class by the former.

\section{Reinforcement Learning}
\label{sec:rl}
Reinforcement Learning (RL) is usually employed to solve a Markov Decision Process (MDP) defined over a real world task. A MDP is defined via a dynamic environment, a state space $\mathcal{S}$, an action space $\mathcal{A}$, and a reward function $R(a,s)$ with $a \in \mathcal{A}$, $s \in \mathcal{S}$~\cite{sutton2018reinforcement}. 
In a MDP, the decision maker, also referred to as \emph{agent}, gets a reward from the environment upon taking an action. The action also causes the environment to change its internal state. The environment is \emph{fully-observable} when the agent observation is always equal to the environment state.
At each time step $t$, the agent receives a state $S_t\in\mathcal{S}$ from the environment, and then selects an action $A_t\in\mathcal{A}$. The environment answers with a numerical reward $R_{t+1}\in\mathcal{R}\subset\mathbb{R}$ and a next state $S_{t+1}$. This interaction gives a \emph{trajectory} of random variables:
\[
S_0,A_0,R_1,S_1,A_1,R_2,\dots
\]
When the agent experiences a trajectory starting at time $t$, it accumulates a \emph{discounted return} $G_t$:
\[
G_t:=R_{t+1}+\gamma R_{t+2}+\gamma^2 R_{t+3}+\dots=\sum_{k=0}^\infty\gamma^k R_{t+k+1},
\]
where the discount factor $\gamma \in [0,1)$ determines the length of the time horizon, so that $\gamma \rightarrow 1$ means an infinite horizon. The return $G_t$ at time $t$ is a random variable, whose probability distribution depends on the policy $\pi(a|s)$, i.e., a deterministic or stochastic strategy that chooses the action $a$ to execute for each possible observation of the environment's state $s$. 
The objective of RL is to find a policy that maximizes the expected discounted reward. 
In this paper, we follow a \emph{deep} reinforcement learning (DRL) approach and use a parametric function approximator for the policy $\pi(a|s)$, specifically a neural network.

\subsection{System Model as a MDP}
\label{sec:mdp}
The application of RL to our task requires to formulate the URLLC scheduling problem as a fully-observable MDP.
Despite the task at hand being inherently not episodic, for convenience of operation we truncate it in multiple episodes of  length $T$ minislots, corresponding to the whole coherence interval of the channel. A minislot $t\in\{1, \dots, T\}$ represents a time step in the episode. At the beginning of each episode, resource allocation and codeword placement for eMMB users is performed and then at each time step, a new URLLC packet is generated with probability $p_u$.

The DRL action consists in deciding whether the first URLLC packet in the queue should be transmitted  in the current minislot or not and on which FR. The \emph{possible actions} at time step $t$ are collected in the set $\mc{A}_t = \{0, 1, \dots, F\}$, where $0$ means no transmission, while otherwise the action indicates the FR index for transmitting the URLLC packet. If the URLLC queue is empty, the only possible action is $0$. 

The \emph{state} at each time step $t$ is then represented by the set $\mathcal{S}_t = \{\mc{S}_t^{(u)}, \mc{S}_t^{(e)}\}$, where $\mc{S}_t^{(u)}$ and $\mc{S}_t^{(e)}$ collect the URLLC and eMBB information at step $t$, respectively.
In particular, the 2-dimensional state $\mc{S}_t^{(u)}$ is 
\begin{equation} \label{eq:stateu}
    \mc{S}_t^{(u)} = \{Q_t, \Delta_t\}
\end{equation}
where $Q_t$ represents the length of the URLLC queue at step $t$, while $\Delta_t = l_u^{\max} - l^\text{old}_t$ represents the difference between the tolerable latency and the latency of the oldest packet in the queue at step $t$. 
The $F$-dimensional state  $\mc{S}_t^{(e)}$  collects for each of the $F$ frequency channels the  variable $s_t(f)$, which tracks if the codeword transmitted on channel $f$ is in outage ($s_t(f)=-1$) or not ($s_t(f)\ge 0$). A non-negative $s_t(f)$ stores the residual number of times that the codeword  can be punctured without being in outage. Let $\rho_{t}(w)$  denote the number of times the codeword $w$ has been punctured from the beginning of the episode, $s_t(f)$ is computed as 
\begin{equation}\label{eq:state:e}
s_t(f) = \max \left\{ C_{w_{t,f}}- \rho_t(w_{t,f}), -1 \right\},
\end{equation}
remembering that $w_{t,f}$ is the codeword placed on resource $f$ and minislot $t$.
Once a codeword is in outage, its state variable is set to -1 and does not change anymore regardless of the times is further punctured.

\subsection{Reward Computation}
\label{sec:reward}
In a RL problem choosing the reward is an empirical process: a good reward function should capture the essence of the task at hand. In this case the objective is to  minimize the number of eMBB codewords in outage while keeping the latency of URLLC packets below the given threshold. With this goal in mind, we introduce the eMMB penalty function $e_t(w)$ 
\begin{equation} \label{eq:rew:e}
e_t(w) =
\begin{cases}
-1,  \quad &\mc{C}_w - \rho_{t-1}(w) \ge 0 \cap \mc{C}_w - \rho_t(w) < 0, \\
0, \quad &\text{otherwise},
\end{cases}
\end{equation}
which takes value  $-1$ only if the  chosen action causes the outage of the  codeword $w$.
Furthermore, since $\Delta_t < 0$ signals the violation of the latency constraint, we introduce the following URLLC penalty function
\begin{equation} \label{eq:rew:latency}
L_t = 
\begin{cases}
0, \quad &\Delta_t \ge 0, \\
-\frac{3 T}{F+1}, \quad &\Delta_t < 0.
\end{cases}
\end{equation}
The heuristic value $-\frac{3 T}{F+1}$ is empirically chosen so that the violation of the latency constraint for an URLLC packet results in a larger negative contribution than the outage penalty for eMBB traffic. 
Accordingly, the reward at time $t$ can be expressed by
\begin{equation} \label{eq:rew}
R_t = \sum_{w \in \mc{W}_t} e_t(w) + L_t,
\end{equation}
Eventually, when $\Delta_t < 0$ the episode is considered finished. 

\subsection{Neural Network (NN) Architecture}
\label{sec:nn}
Among the possible RL techniques, we consider a DRL Policy Gradient (PG) algorithm called Proximal Policy Optimization (PPO)~\cite{schulman2017proximal}. PPO aims at taking the biggest possible improvement step on a policy without ending too far from the starting point one, thus avoiding the risk of performance collapse. PPO is an actor-critic algorithm~\cite{schulman2017proximal}, where two different neural networks are required. To this respect, we consider two completely separated subnetworks, one for the \emph{value function} (estimated value of the state) and one for the \emph{policy function} (the strategy).
Both policy and value function subnetworks have three dense layers with 128, 64, and 32 neurons, respectively. All of them operate a rectified linear activation function (ReLU). Furthermore, the policy subnetwork has a dense fourth layer with $F + 1$ neurons to choose the actions, while the value subnetwork has a dense fourth layer with $1$ neuron and no activation  to estimate the value. Finally, all layers are initialized using Xavier initialization. For additional details refer to the GitHub repository~\cite{telerl2021github}.

\subsection{Framework}
\label{sec:framework}

The RL framework adopted in this study is USienaRL\footnote{Available on PyPi and also on GitHub: \url{https://github.com/InsaneMonster/USienaRL}.}. This framework allows for environment, agent and interface definition using a preset of configurable models. While agents and environments are direct implementations of what is described in the RL theory, interfaces are specific to this implementation. Under this framework, an interface is a system used to convert environment states to agent observations, and to encode agent actions into the environment. This allows to define agents operating on different spaces while keeping the same environment. By default an interface is defined as pass-through, i.e.\ a fully observable state where agents action have direct effect on the environment. We always use pass-through interfaces through this work. Moreover, within this framework, the comparison schemes described in Section~\ref{sec:results} are also defined as agents operating without a neural model, using only a set of predefined rules acting w.r.t.\ the states of the environment.

\section{Results}
\label{sec:results}
We consider a simplified scenario, where the slot duration and the coherence time of the channel  are set to $1$ and $10$ ms, respectively. Each slot is further divided in  $M=14$ minislots. The number of frequency resources is $F=12$. The length of an episode corresponds to the coherence time of the channel so that the number of time slots for each episode is $\Sigma = 10$, for a total of $T = 140$  minislots. We consider a single URLLC user, i.e. $|\mc{U}| = 1$, and the number of eMBB users is $|\mc{E}| = 10$. We further set the maximum delay constraint to $l_u^{\max} = M/2 = 7=0.5$ ms. 
We consider only codewords of class $C_w \in \left\{0,1\right\}$, i.e., codewords that can be punctured zero or one times before being in outage. 

Regarding PPO, we use an instance of PPO-Clip as described by OpenAI at~\cite{openai2018ppo}, with differentiated value and policy heads. We use a clip ratio equal to $0.2$, and an early stopping strategy if the mean KL-divergence of the new policy from the old one grows beyond a given threshold. More specifically, we set as threshold the value $1.5\cdot 10^{-2}$. 
To reduce the variance, we use the generalized advantage estimation approach as proposed in~\cite{schulman2015high}, with $\gamma_{\rm{GAE}} = 1$ and $\lambda_{\rm{GAE}} = 0.97$.

To have a fair performance comparison, we consider the  three alternative URLLC scheduling algorithms:
\begin{itemize}
    \item \emph{Random}. The decision whether to transmit or not the URLLC packet is randomly taken with equal probability. In case of transmission, the frequency is selected randomly with uniform probability distribution. 
    \item \emph{Aggressive}. The URLLC packet is transmitted immediately on a randomly chosen frequency.
    \item \emph{Threshold Proportional (TP)}. The URLLC packet is transmitted immediately on the frequency resource occupied by the codeword with the highest puncturing threshold, given by~\eqref{eq:state:e}. TP has almost optimal performance when the URLLC is forced to transmit immediately upon arrival, i.e., $l_u^{\max} = 1$, and in case of low average URLLC load~\cite{Anand2020}.
    \item  \emph{TP-lazy}. As long as $\Delta_t > 0$, the packet is transmitted only if $\sum_{w\in\mc{W}_t} C_w - \rho_t(w) \ge \sum_{w\in\mc{W}_{t+1}} C_w - \rho_t(w)$, i.e. if the present state is somehow better (or equal) than the next one. If $\Delta_t = 0$, the transmission is forced in the present minislot. In any case, the choice of the frequency is made according to the TP scheme. This heuristic combines the advantage of the TP transmission policy with the possibility of waiting before puncturing eMBB resources. 
\end{itemize}
During the \emph{learning phase} of the PPO agent, the parameters related to eMMB resource allocation and URLLC traffic generation are randomized on an episode basis. 
While this is not mandatory to train a functioning agent, it is crucial to help the agent to generalize  the task at hand. In other words, the agent is requested to learn a generalized strategy that is not specific either for a particular eMMB allocation policy or a particular URLLC traffic load.       
Hence, the class of each codeword and the probability of generating a URLLC packet $p_u$ is randomly chosen. 
Regarding the codeword placement, the resource allocation of each eMBB user is performed randomly in the resource grid; then, imposing $|\mc{W}| = 120$, the codeword are generated  with random length of $|w|$ minislots, in order to occupy the whole resource grid. For sake of simplicity, we assume that each codeword transport the same quantity of information regardless of its length. This assumption will be relaxed in future works.
Regarding the classes of the different codewords, we limit our simulation analysis to considering $C_w \in \left\{0,1\right\}$, i.e., no puncturing or puncturing of a single minislot is allowed. As for the case of resource allocation, also the class assignment is random. In particular, at the beginning of each episode, the environment select randomly one of the following five possible distributions:
\begin{equation} \label{eq:classes}
D \in \{[0, 1], [0.2, 0.8], [0.5, 0.5], [0.8, 0.2], [1, 0]\}
\end{equation}
where the first and the second element of each vector are the percentage of the codeword of class 0 and class 1, respectively. More specifically, if $D = [0.5, 0.5]$, then 50\% of the placed codewords have $C_w = 0$ and the other 50\% have $C_w = 1$, and so forth. Furthermore, the URLLC traffic load $p_u$ is also randomized in each episode. In particular, we consider:
\[
p_u \in \{0.1, 0.2, 0.3, 0.4, 0.5\}
\]
where the value is chosen according to a uniform probability distribution.
Since the MDP formulation considered in this paper is not inherently episodic, for the computation of the discounted return we set $\gamma=0.99$.

To simulate a continuous task, we initialize each episode with a random number of URLLC packets in the queue. The number of packets generated in this way is always smaller than $l_u^{\max}$ to avoid that the episode starts with $\Delta_t < 0$.

After that the NN has been trained, we can show the results obtained by running the RL-based agent in \emph{inference mode} and provide comparisons with the considered heuristic alternatives. 

In all simulations, all schemes \emph{always satisfy the URLLC latency constraints}: aggressive, TP, and TP-lazy by design, PPO because of the choice of a proper reward function that greatly penalizes the loss of a URLLC packet.

In the following we show the results obtained by running a simulation after training the RL-based agent, i.e.\ in inference mode, and we provide comparisons with the considered heuristic alternatives. 
More specifically, the results are collected over 5 simulation runs of 1000 episodes each. In each simulation run we consider a different value of the activation probability $p_u$ ranging in the set $\{0.1, 0.2, 0.3, 0.4, 0.5\}$. At each episode, the codeword placement probability vector $D$ is randomly selected in the set $\{[0, 1], [0.2, 0.8], [0.5, 0.5], [0.8, 0.2], [1, 0]\}$.
First, we present the results obtained with $T = 140$, as in the training phase, in order to show that the PPO agent is able to learn an optimal policy. Then, we will show that the PPO agent trained with the episodic behaviour can be applied even for continuous tasks.

\begin{figure}[htb]
    \centering
    \includegraphics[width=10cm]{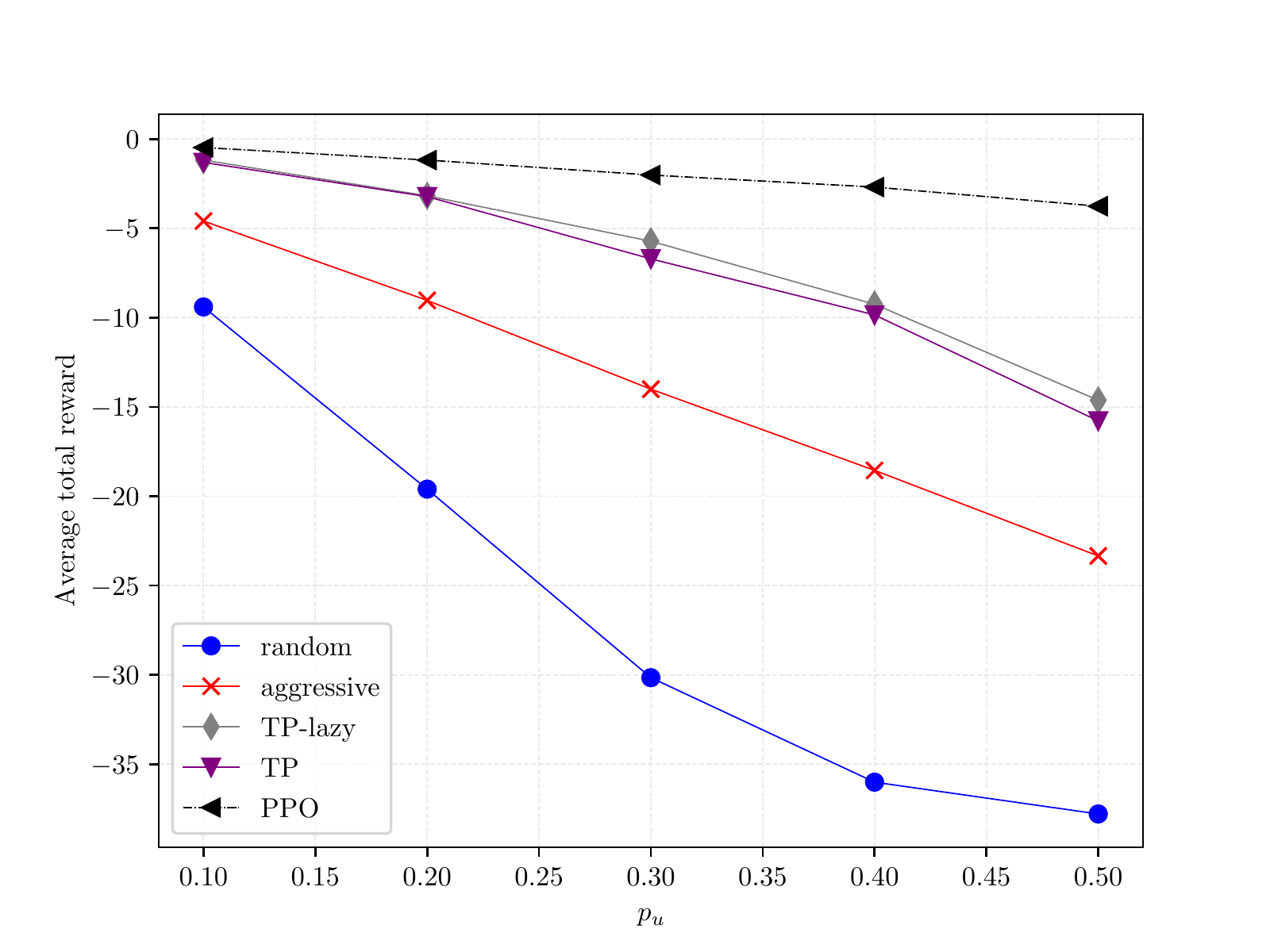}
    \caption{Total reward versus activation probability $p_u$.}
    \label{fig:tot_avg_reward}
\end{figure}

Figure~\ref{fig:tot_avg_reward} shows the total episode reward $\sum_{t=1}^T R_t$ as a function of $p_u$, for $T=140$. It is worth noting that the PPO agent, trained with random values of $p_u$, outperforms all the other schemes for every value of $p_u$, thus assessing the generalization capability of the network. 
In Table~\ref{tab:residual}, we further show the average number of packets remaining in the URLLC queue at the end of an episode for different $p_u$. The results for TP and aggressive are omitted since in both cases URLLC packets are promptly transmitted upon arrival. The PPO agent learns a policy that attempts to keep the URLLC queue almost empty since the opposite could lead to a violation of latency constraint. Conversely, in the TP-lazy case, where no control is operated onto the queue length, a non-negligible amount of traffic remains unserved at the end of an episode.

\begin{table}[htb]
    \centering
    \begin{tabular}{cccccc}
        \toprule
        $p_u$ & 0.1 & 0.2 & 0.3 & 0.4 & 0.5 \\
        \midrule
        TP-lazy & 0.597 & 1.222 & 1.790 & 2.416 & 3.015 \\
        PPO & 0.030 & 0.068 & 0.081 & 0.136 & 0.210 \\
        \bottomrule
    \end{tabular}
    \vspace*{2mm}
    \caption{Average number of URLLC packets not served before the end of the episode.}
    \label{tab:residual}
\end{table}

The subdivision of the task into episodes of a fixed length may somehow distort the correct evaluation of the algorithms' performance. To simulate a longer time horizon is of critical importance because the agent should be able to correctly work without the artificial partitioning of the task into episodes. To address this issue, we scaled up the length of each episode by one order of magnitude, \emph{without retraining the agent}.

\begin{figure}
    \centering
    \includegraphics[width=10cm]{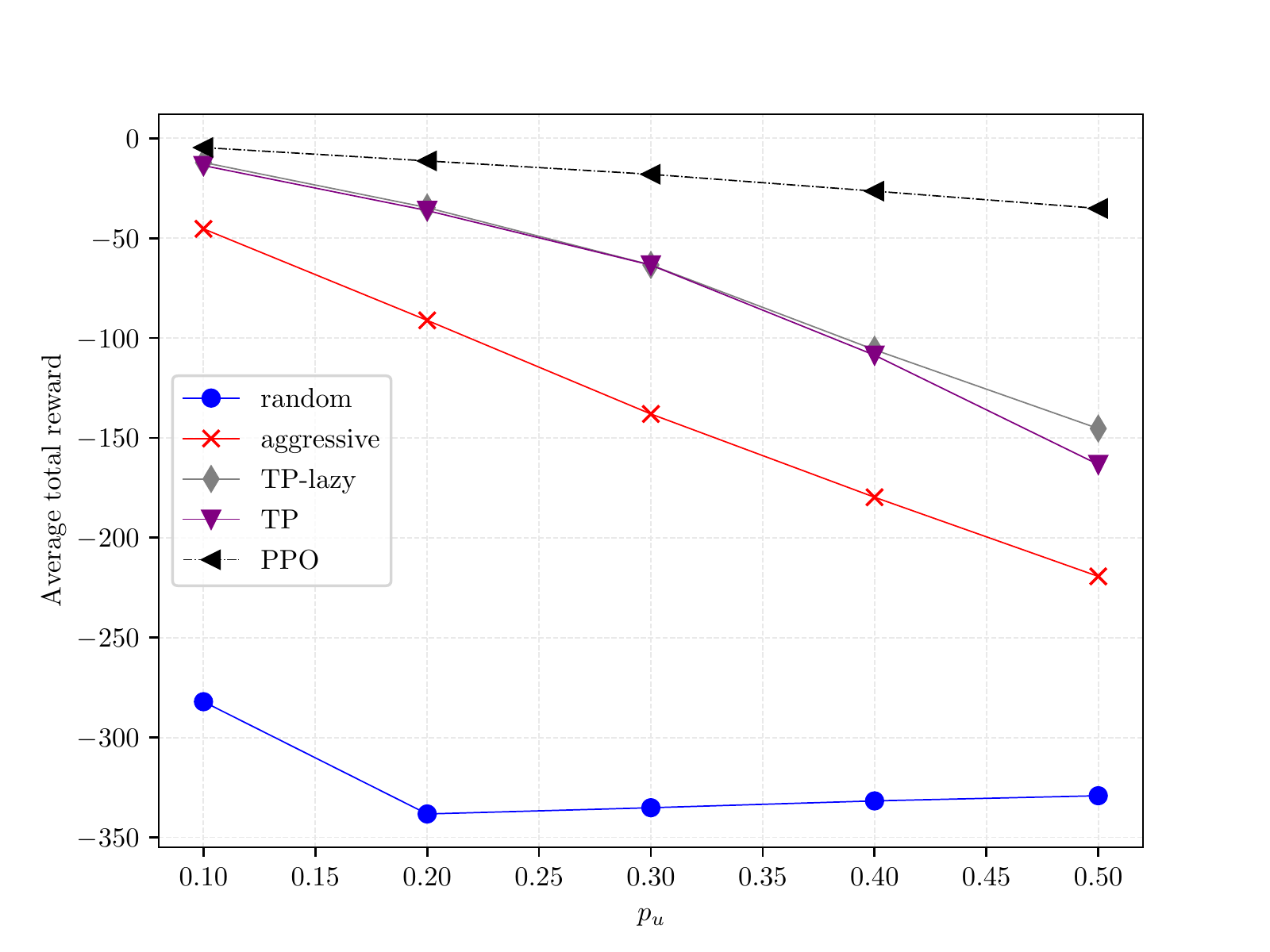}
    \caption{Average total reward versus $p_u$ with $T = 1400$.} 
    \label{fig:T}
\end{figure}

Figure~\ref{fig:T} shows the performance of the RL agent as a function of $p_u$ with the length of each episode increased by one order of magnitude. Specifically, the results are obtained for $T = 1400$, increasing also $|\mc{W}| = 1200$ for coherence; everything else is left the same.  We can see that the RL agent is still able to outperforms all the heuristics for each value of $p_u$, thus proving the learned policy to be independent from the artificial subdivision of the task in episodes. 

\begin{figure}[htb]
    \centering
    \includegraphics[width=10cm]{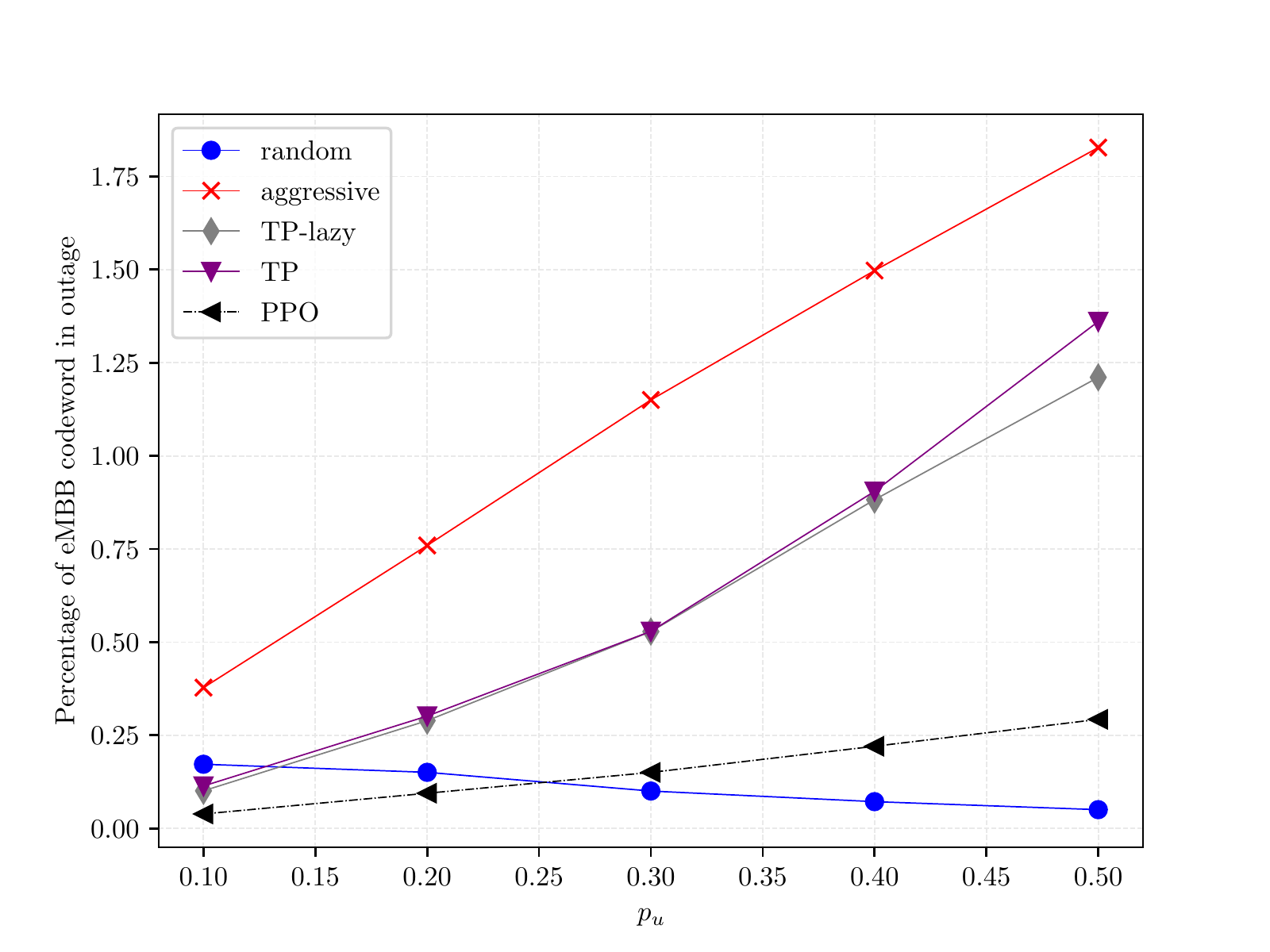}
    \caption{Percentage of eMBB codeword in outage versus activation probability $p_u$, $T = 1400$.}
    \label{fig:embb_outages}
\end{figure}

Figure~\ref{fig:embb_outages} shows the percentage of eMMB codewords in outage at the end of each episode for the different schemes, for $T = 1400$, while the class of each codeword is again randomly chosen. It is clear that the PPO approach is able to outperform all the other schemes except the random one, for all the considered $p_u$. 
It is worth noting that the random scheme may achieve better performance for high only because most episodes are stopped due to the latency violation, which occurs often, as shown in Table~\ref{tab:urllc_delays}.
The other schemes never violate the latency constraint, and thus they are not presented in the Table.

\begin{table}[htb]
    \centering
    \begin{tabular}{c|ccccc}
        $p_u$ & 0.1 & 0.2 & 0.3 & 0.4 & 0.5 \\
        \toprule
        $\Pr(\Delta_u < 0)$ & 0.166 & 0.379 & 0.674 & 0.883 & 0.982 \\
        \bottomrule
    \end{tabular}
    \vspace*{2mm}
    \caption{Probability of latency constraint violation of random scheme.}
    \label{tab:urllc_delays}
\end{table}

\begin{figure}
    \centering
    \includegraphics[width=10cm]{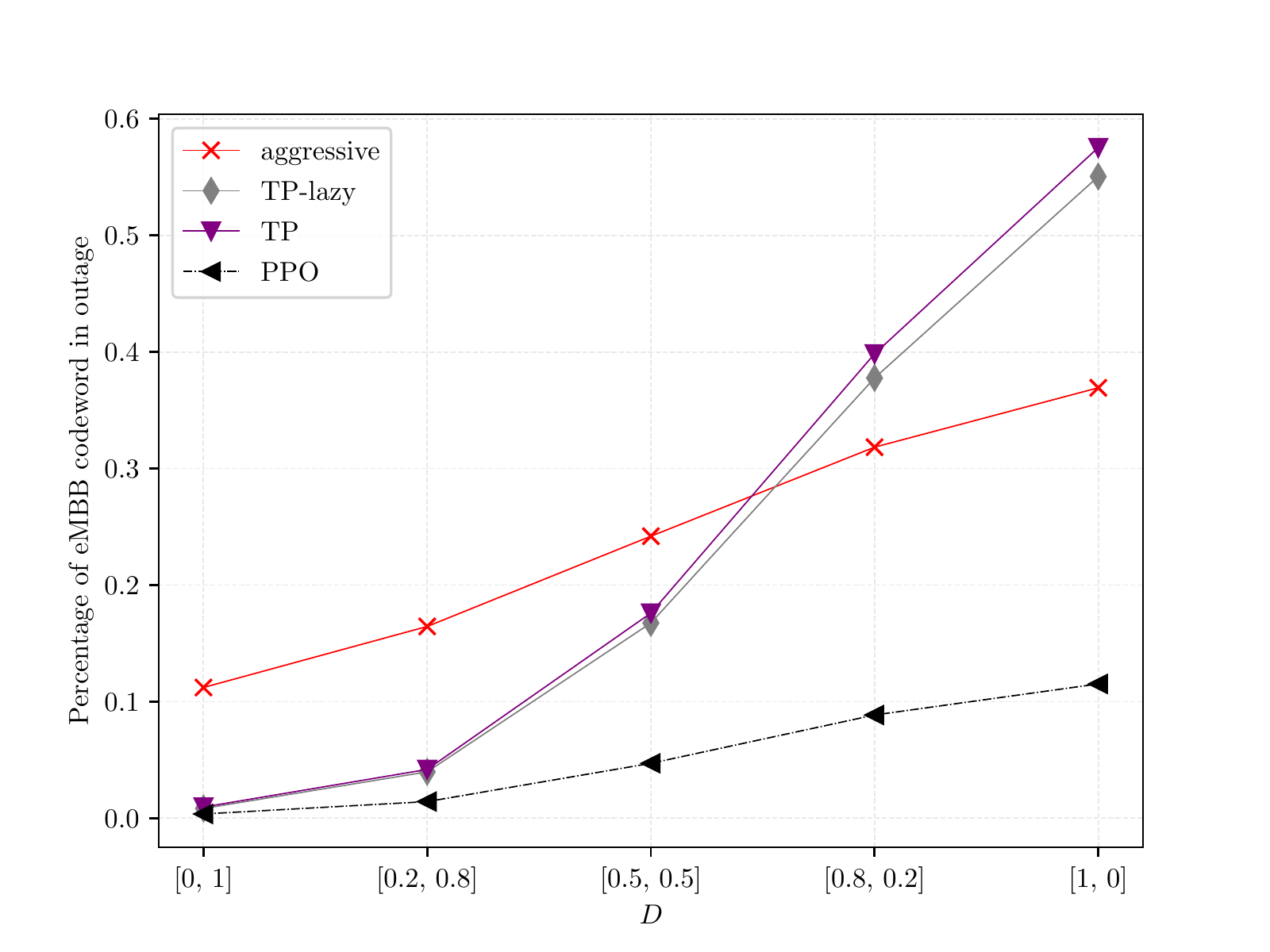}
    \caption{Percentage of eMBB codewords in outage versus the different percentage of classes of codeword for $p_u = 0.5$.}
    \label{fig:varyingD}
\end{figure}

Fig.~\ref{fig:varyingD} shows the percentage of eMBB codewords in outage for different compositions of eMBB codewords classes $D=[\Pr\{\mc{C}_0\}, \Pr\{\mc{C}_1\}]$ and $p_u = 0.5$.
The PPO agent outperforms by a wide margin all other schemes, showing the versatility of the RL approach. Among the other things, these results show that PPO has good performance even when $D=[1, 0]$ and there are only codewords without an inner erasure code.

\section{Conclusions}
\label{sec:conclusions}
We proposed a deep reinforcement learning approach based on PPO, which is able to dynamically manage the coexistence of the URLLC traffic on top of the eMBB traffic.
The trained RL agent overall outperforms all the other schemes on multiple performance metrics, being capable of noteworthy generalization over different tasks. 
Our approach is highly scalable with respect to the length of each simulation, without retraining the agent. This is of critical importance since the real world task we modeled is inherently not episodic and the artificial subdivision in episodes is only required to train an agent in an RL fashion.

We believe this work to be a promising step into the direction of solving the task of eMBB-URLLC resource slicing. Our future works will mainly focus on:
\begin{itemize}
    \item taking into account the reliability of URLLC user;
    \item adopting a Poissonian distribution to simulate the arrival of URLLC packets;
    \item addressing the transmission over multiple frequency resources;
    \item enabling non-orthogonal multiple access communication by means of superposition coding at the transmitter and successive interference cancellation at receivers.
\end{itemize}

\clearpage

\printbibliography

\end{document}